\documentclass[aps,prl,twocolumn,showpacs]{revtex4}
\usepackage{epsfig}
\usepackage{graphicx}
\usepackage{amsfonts}
\usepackage[figuresright]{rotating}
\usepackage{amssymb}
\usepackage{amsmath}
\usepackage{dcolumn}
\usepackage{bm}

\def\be{\begin{equation}} \def\ee{\end{equation}}
\def\bea{\begin{eqnarray}} \def\eea{\end{eqnarray}}

\begin{document}
\title{Mott made easy}
\author{Congjun Wu}
\affiliation{Department of Physics, University of California,
San Diego, CA92093}
\begin{abstract}
The realization of a Mott insulating state in a system of ultracold fermions 
comprising far more internal components than the electron, provides an 
avenue for probing many-body physics that is difficult to access in solids.
\end{abstract}
\maketitle

Particle physicist Isaak Pomeranchuk's proposal \cite{pomeranchuk1950} 
in 1950 that $^3$He 
might be cooled simply by squeezing it was based on a striking, if 
straightforward, observation: Pomeranchuk saw that the entropy of solid $^3$He 
exceeded that of liquid $^3$He at very low temperatures, and inferred that 
compressing a mixture of the two would result in cooling, as the liquid was 
converted into solid \cite{richardson1997}. Writing in Nature Physics, 
Shintaro Taie and colleagues have made use of this idea to achieve a 
Mott-insulating state in $^{173}$Yb — revealing an important distinction 
between the symmetry classes of ultracold-atomic ensembles \cite{taie2012}.

Mott insulators are materials that are expected to be metallic according 
to band theory, but are instead driven by strong repulsions between 
electrons into an insulating state. 
Compressing a Mott insulator comprising one electron per lattice site 
creates charge excitations of doubly occupied sites, on which two 
electrons repel each other more strongly than they would when apart. 
At low enough temperatures, these charge excitations are frozen and so 
compressibility is suppressed. 
In the experiment performed by Taie et al., $^{173}$Yb fermions were loaded 
into optical lattices, and their Mott insulating behaviour was confirmed by 
demonstrating this suppression of compressibility, together with the
appearance of a charge-excitation energy gap.

Unlike electrons in solids, many fermionic atoms carry large hyperfine spins, 
which are predicted to exhibit exotic states with high symmetries that 
usually only appear in high-energy physics \cite{wu2003,lecheminant2005,wu2006,
gorshkov2010}. 
In the case of $^{173}$Yb atoms, the hyperfine spin is $F = \frac{5}{2}$, 
meaning that the number of spin components is $N = 2F + 1 = 6$. 
Because the electronic shells of $^{173}$Yb atoms are fully filled, their 
hyperfine spins are simply nuclear spins. 
And as nuclei are buried deep inside atoms, all nuclear-spin components 
interact in the same way. The $^{173}$Yb system is therefore invariant 
under any unitary transformation within this six-dimensional hyperfine-spin 
space — affording it the symmetries associated with the SU($N = 6$) group. 
By contrast, two-component electron systems usually possess SU($N = 2$) 
symmetry. 
Aside from their applications in high-energy physics, the symmetries of 
SU($N \ge 3$) have been used to study quantum antiferromagnetism, 
albeit only as a mathematical tool. 
Now, with the work by Taie et al., experimental explorations of these highly 
symmetric Mott insulators have really begun.

Taie et al. found cooling their multicomponent $^{173}$Yb system towards 
the Mott insulating state to be more effective \cite{taie2012} than the more widely used 
two-component systems \cite{jordens2010}. In fact, they showed that driving 
fermions adiabatically into a Mott insulating state can lead to a type 
of cooling analogous to Pomeranchuk cooling -- an idea corroborated by 
simulation studies \cite{hazzard2012,cai2012}.

The analogy is based on the fact that fermions can hold more entropy in 
a Mott insulating state than in a Fermi-liquid state. 
This is because every site in a Mott insulating state contributes to entropy 
through spin configurations, unlike in a Fermi liquid, in which only fermions 
close to Fermi surfaces contribute. 
The Mott insulating state realized in the experiment by Taie et al. 
has one atom per site. 
Because the entropy per site scales as $\ln N$, as $N$ increases, 
the system can accommodate more entropy. 
This in turn leads to a significant reduction in the temperature at which 
the Mott-insulating state can be achieved — effectively cooling the 
system through transfer of entropy from orbital motions to spin configurations in a manner reminiscent of Pomeranchuk cooling. 
This cooling results in a less compressible gas, invoking the signature of the Mott insulator.

One of the key features associated with the SU(6) symmetry realized by 
Taie et al.3 is the strong quantum hyperfine-spin fluctuations enabled by 
large values of $N$ \cite{wu2006}. 
This may seem to contradict the standard lore in solids that quantum fluctuations of large spins are weak, but the reasoning is straightforward. 
Large spin can build up in solids, as Hund's rule dictates, if the electron 
shells of cations are partially filled. Although the value of spin $S$ may be large, its symmetry remains SU(2). 
Quantum fluctuations, which arise from the non-commutativity between three 
spin components, become weaker as $S$ increases. By contrast, the $^{173}$Yb 
system enjoys a much higher symmetry than these electron systems, and thus 
the hyperfine spins fluctuate in a much larger internal phase space — 
meaning that large values of $N$ enhance, rather than suppress, these 
fluctuations. 
Quantum fluctuations in the $^{173}$Yb system are actually even stronger
than those of spin-$\frac{1}{2}$ electrons.

These strong quantum fluctuations enhance the tendency of the SU($N \ge 3$) 
antiferromagnets to form singlets. 
Just like quantum chromodynamics, in which the SU(3) colour singlet states 
of baryons consist of three quarks, the minimum site number required to 
form an SU($N$) singlet is just $N$. 
This means that in the $^{173}$Yb system probed in the experiment by 
Taie et al.3, quantum-antiferromagnetic fluctuations are dominated by 
six-site correlations \cite{wu2006,hermele2011}, whose physics cannot be 
reduced into two-site correlations as in the extensively studied case of 
SU(2) quantum magnets in solids \cite{jordens2010}.

The experiment performed successfully by Taie et al.3 provides a new 
opportunity to study novel Mott insulators that are difficult to realize 
in solids. 
Indeed, the reduction in temperature of an SU(6) gas relative to an SU(2) 
gas may prove essential to our realization of exotic spin order in these 
systems. 
However, in order to study SU(6) quantum antiferromagnetism, for example, 
further cooling is necessary. 
Although it is still beyond the current experimental capability, one hopes 
that techniques for sufficient cooling will be achieved in the near future. 
These advances will not only enhance our understanding of the Mott insulating 
state, but may also enable realization of exotic spin-liquid states
\cite{hermele2011,cai2012b}
 and fermionic superfluid states induced by doping these unusual Mott insulators.

\end{document}